\documentstyle[prl,aps,twocolumn,psfig]{revtex}

\begin{document}

\title{How to observe the Efimov effect}
\author{E.~Nielsen, D.V.~Fedorov and A.S.~Jensen}
\address{ Institute of Physics and Astronomy,
Aarhus University, DK-8000 Aarhus C, Denmark}

\twocolumn
\maketitle
\draft

\begin{abstract}
We propose to observe the Efimov effect experimentally by applying an
external electric field on atomic three-body systems.  We first derive
the lowest order effective two-body interaction for two spin zero atoms
in the field. Then we solve the three-body problem and search for the
extreme spatially extended Efimov states. We use helium trimers as an
illustrative numerical example and estimate the necessary field strength
to be less than 2.7 V/{\AA }.
\end{abstract}


\paragraph*{Introduction.} 

The Efimov effect \cite{efi70} is an interesting anomaly in a three-body
system with short-range interactions. If at least two of the binary
subsystems have an $s$-state at zero energy (or equivalently an infinitely
large scattering length) the three-body system develops an infinite series
of bound states.  The binding energies of these states are exponentially
small and the spatial extension is correspondingly exponentially large
\cite{fed93}.

A peculiar feature of these states is that their number decreases when
the modulus of the scattering length is decreased by either weakening
or {\em strengthening} the two-body potentials.  For a given scattering
length $a$ and effective range $r_e$ the number of bound three-body states
is proportional to $\log{(|a|/r_e)}$.  This number becomes infinitely
large also when the effective range approaches zero, which in this limit
is called the Thomas effect~\cite{tho35}.

Although the Thomas effect  has  only  theoretical  interest, the Efimov
effect might be observed experimentally in a suitable weakly bound
three-body system.  Candidates for the Efimov effect are Borromean
(that is without bound subsystems) halo nuclei, occurring at the
neutron drip-line \cite{fed94}, and molecular systems like the atomic
helium trimers \cite{esr96,mot98,nie98}.  For the $^4$He-trimer two
bound states are predicted with the excited state resembling an Efimov
state.  The ratio $|a|/r_e$ is in this case around 14. The excited state
disappears when the interaction is either weakened by 3\% or strengthened
by 20\%.  So far other examples are not available since the condition
of the exceedingly large scattering length is hard to meet.

This difficulty could be overcome if one could gradually alter the
two-body potentials, thus adjusting the scattering length to the needed
condition.  An already large scattering length is very sensitive to
small variations of the potential and the needed alteration should
therefore be relatively small. Such tuning of the scattering length in
an external magnetic field was recently suggested for a system of two
rubidium atoms \cite{vog97}. A corresponding Feshbach resonance at zero
energy was subsequently observed experimentally \cite{cou98}. Another
similar resonance has also been observed for a system of two sodium atoms
\cite{ino98}.  However corresponding investigations of three-body systems
do not exist.  We have therefore undertaken the first investigation of
a three-body system in an external field with the main emphasis on the
Efimov effect.

So far the most promising candidate for the Efimov effect is believed
to be the system of three helium atoms. For the noble atoms, unlike
the alkali atoms, the electric rather than magnetic field seems to be
most suitable instrument.  In this letter we suggest to use an external
electric field as a tool to modify the interaction between noble atoms.
We shall consider examples of helium isotopes and estimate the electric
field strength needed to reach the Efimov limit.  We shall investigate the
problem of three helium atoms in an external electric field and estimate
the properties of the lowest Efimov states. We shall also discuss the
feasibility of an experiment observing the Efimov effect in atomic or
molecular three-body systems.

\paragraph*{Three-body system in an external field.}

An external field breaks the rotational symmetry and a precise
computation of the Efimov states is exceedingly difficult. It is
already difficult for systems with well defined angular momentum
\cite{fed94}.  We shall therefore assume that our binary subsystems
are already close to the threshold and that the needed external field
is weak enough to justify a perturbative treatment of the problem.

We can not, however, apply the perturbation theory directly to the
three-body problem since notwithstanding its weakness, this perturbation
supposedly leads to a significant modification of the spectrum --
appearance of an infinite series of bound states.

On the other hand on the two-body level this perturbation only slightly
modifies the binary interactions.  Our strategy is therefore to estimate
perturbatively the variation of the potential between atoms exposed
to an external field and then solve the three-body problem with the
modified potentials.

\paragraph*{Correction to the two-body potential.}

We shall now calculate the total energy of the system
of two atoms exposed to a weak static external electric field.  One of the
terms in this energy is, within the usual adiabatic approximation,
the sought correction to the potential.

The Hamiltonian of a system of two atoms separated by a fixed distance
${\bf r}$ in an external static electric field ${\bf\cal E}$ can be
written in the dipole approximation as
\begin{eqnarray}
H&=&H_0^{(1)}+H_0^{(2)}+\Delta H \nonumber\\
\Delta H &=& - {\bf d}^{(1)}\cdot{\bf\cal E}- {\bf d}^{(2)}\cdot{\bf\cal E}\\
&+& {{\bf d}^{(1)}\cdot{\bf d}^{(2)} -
 3({\bf d}^{(1)}\cdot\hat{\bf r})({\bf d}^{(2)}\cdot\hat{\bf r})\over r^3} \; , \nonumber
\end{eqnarray}
where $H_0^{(i)}$ and ${\bf d}^{(i)}$ respectively are the unperturbed
Hamiltonian and the dipole operator of the atom $i$ ($i=1,2$).  The
unperturbed state of the atom $i$ with the principal quantum number
$n_i$, angular momentum $l_i$ and its projection $m_i$ on the
direction of $\hat{\bf r}\equiv{\bf r} /|{\bf r}|$ is denoted by
$|n_il_im_i\rangle$. The corresponding energy is $E_{n_i}$ with the
ground state energy set by definition to zero.  As a shorthand
notation we shall use $\nu_i = n_il_im_i$ and
$|\nu_1\nu_2\rangle=|\nu_1\rangle|\nu_2\rangle$.  The ground states
will be denoted as $|0_i\rangle$ and $|0\rangle =
|0_1\rangle|0_2\rangle$.  

The operators ${\bf d}^{(i)}$ have negative parity and therefore the
first order correction to the energy $\Delta E^{(1)}=\langle 0|\Delta H
|0\rangle$ is zero since $\left<0_i|{\bf d}^{(i)}|0_i\right>=0$. The
second order correction

\begin{equation}  \label{sec_order}
\Delta E^{(2)}=  - \sum_{\nu_1\nu_2}
\frac{\langle 0 | \Delta H |\nu_1 \nu_2 \rangle
\left<\nu_1 \nu_2 |\Delta H |0\right>}
{E_{n_1}+E_{n_2}} \; 
\end{equation} 
can be rewritten in terms of the reduced dipole matrix element
$d_{n_i}$ defined by
\begin{equation} \label{d_def}
\langle n_il_im_i|{\bf d}^{(i)}\cdot{\bf\cal E}|0_i\rangle
\equiv\delta_{l_i,1}d_{n_i}{\bf\cal E}_{m_i} \; ,
\end{equation}
where ${\bf\cal E}_{m}$ ($m=0,\pm 1$) is the usual spherical tensor component
of the vector ${\bf\cal E}$. By using this in Eq.(\ref{sec_order}) we get
\begin{eqnarray} 
\Delta E^{(2)}=
-{\sum_{\nu_1}}'{|d_{n_1}{\bf\cal E}_{m_1}|^2\over E_{n_1}}
-{\sum_{\nu_2}}'{|d_{n_2}{\bf\cal E}_{m_2}|^2\over E_{n_2}}  \nonumber\\
-{1\over r^6}{\sum_{\nu_1\nu_2}}'
 {|d_{n_1}d_{n_2}(1-3\delta_{m_1,0})|^2
\over E_{n_1}+E_{n_2}}\delta_{m_1,-m_2}
 \; ,
\label{second_order}
\end{eqnarray}
where the primed summation sign indicates that only states with $l_i=1$
are included.  The first two terms are the usual polarization terms
$-{1\over 2}\beta_i{\bf\cal E}^2$, where the polarizability $\beta_i$
in our notation is given by \cite{han91}
\begin{equation}
\beta_i=2{\sum_{n_i}}'{|d_{n_i}|^2\over E_{n_i}} \; .
\end{equation}
These terms can be effectively used to guide single atoms in
electrostatic lenses \cite{ket92}. However they do not change the
interaction between atoms as they only give rise to a constant shift
of the total energy.  The last term in Eq.(\ref{second_order}) is the
long-range dipole-dipole part of the bare Van der Waals interaction
without the external field.

The lowest order correction, caused by the field ${\bf\cal E}$, to the
two-body interaction appears in the third order perturbation term
\begin{eqnarray}
\Delta E^{(3)}=\sum_{\nu_1\nu_2}\sum_{\nu'_1\nu'_2}
(E_{n_1}+E_{n_2})^{-1}(E_{n'_1}+E_{n'_2})^{-1}\nonumber\\
\times
\langle 0           | \Delta H |\nu_1\nu_2  \rangle
\langle \nu_1\nu_2  | \Delta H |\nu'_1\nu'_2\rangle
\langle \nu'_1\nu'_2| \Delta H |0           \rangle 
\; .  \label{E3_gen}
\end{eqnarray}
Using the definition in Eq.(\ref{d_def}) we then get
\begin{eqnarray}
\Delta E^{(3)}={4\over r^3}{\sum_{\nu_1\nu_2}}'
{ |d_{n_1}|^2|d_{n_2}|^2 \over E_{n_1}E_{n_2} }\nonumber\\
\times{\bf\cal E}_{m_1}{\bf\cal E}_{m_2}\left(1-3\delta_{m_1,0}\right)\delta_{m_1,-m_2}
 \nonumber\\
=\beta_1\beta_2 |{\bf\cal E}|^2 {1-3\cos^2\theta\over r^3}
=-\beta_1\beta_2 |{\bf\cal E}|^2 \sqrt{16\pi\over 5}
{Y_{20}(\theta)\over r^3} \; ,
\label{e3}
\end{eqnarray}
where $\theta$ is the angle between ${\bf r}$ and ${\bf\cal E}$.  This term
is clearly the interaction between two classical dipoles
${\bf D}^{(i)}=\beta_i{\bf\cal E}$ induced by the field on the atoms.

Higher order corrections have higher powers in either field or
polarizability and they are presumably smaller.  However, the fourth
order correction, with the third power of polarizability, includes a
possibly important term which is proportional to $|{\bf\cal E}|^2$,
similarly to $\Delta E^{(3)}$.  We shall estimate this small
term approximately by use of the semiclassical operator
\begin{eqnarray}
\delta H=
{{\bf D}^{(1)}\cdot{\bf d}^{(2)} 
      - 3({\bf D}^{(1)}\cdot\hat{\bf r})
	({\bf d}^{(2)}\cdot\hat{\bf r})\over r^3}
 \nonumber\\
+{{\bf d}^{(1)}\cdot{\bf D}^{(2)} 
      - 3({\bf d}^{(1)}\cdot\hat{\bf r})
	({\bf D}^{(2)}\cdot\hat{\bf r})\over r^3} \; .
\label{dh}
\end{eqnarray}
Each term in this operator corresponds to the interaction of an atom
with a classical dipole $\bf D$.

The second order perturbation theory now gives for the first term in
Eq.(\ref{dh})
\begin{eqnarray}
\delta E_1=
-{\sum_{\nu_2}}'{|d_{n_2}(1-3\delta_{m_2,0})D^{(1)}_{m_2}|^2
\over E_{n_2} r^6 }
\nonumber\\
=-{1\over 2} \beta_2 |{\bf d}^{(1)}|^2{1+3\cos^2\theta\over r^6}\nonumber\\
=-\beta_1^2\beta_2 |{\bf\cal E}|^2 \sqrt{4\pi} 
{Y_{00}(\theta)+{1\over\sqrt{5}}Y_{20}(\theta)\over r^6} \; . \label{de}
\end{eqnarray} 
The contribution $\delta E_2$ from the second term in Eq.(\ref{dh}) is
directly obtained from Eq.(\ref{de}) by interchanging the atomic indices
1 and 2.

In total we shall use the following correction to the
two-body potential due to the external field
\begin{eqnarray}
\Delta V(r)=\Delta E^{(3)}+\delta E_1+\delta E_2 \nonumber \\
=-\beta_1\beta_2\sqrt{16\pi\over 5}{Y_{20}(\theta)\over r^3}|{\bf\cal E}|^2
 \nonumber\\
 -\beta_1\beta_2(\beta_1+\beta_2)\sqrt{4\pi}
{Y_{00}(\theta)+{1\over\sqrt{5}}Y_{20}(\theta)\over
r^6}|{\bf\cal E}|^2 \; . \label{dv}
\end{eqnarray}
Both terms in this two-body interaction are of second order in the
external field. They differ in the radial dependence and the order of
the polarizability.  The extra polarizability factor and the faster
radial fall off of the second term produce a relatively smaller
contribution.  The terms with $Y_{20}$ break the rotational symmetry
and couple different angular momenta.  The individual properties of
the two-body system determine the relative contribution of the two
terms.  Since both terms resulted from the second order perturbation
treatment they must lower the ground state energy of the two-body
system.

\paragraph*{Numerical examples.}
To observe the Efimov effect introduced by an external field we have to
select a three-body system where at least two of the binary subsystems
are almost bound in relative $s$-states and the third subsystem is
unbound. The atomic helium trimers seem to offer an interesting and
realistic possibility where the necessary properties are present and
accurately known, i.e. $^4$He$_2$ is weakly bound and $^3$He$^4$He is
marginally unbound \cite{esr96,mot98,nie98}. Thus the $^4$He$^3$He$_2$
system is a realistic candidate where the external field then may provide
the additional energy needed to bind $^3$He$^4$He.  It would probably
be even more advantageous to aim for three identical bosons.

We use the local and central potential LM2M2 as the free-space
interaction between He atoms \cite{azi93} and the polarizability is
$\beta=0.2050\times 10^{-24}$cm$^3=1.383$ a.u. \cite{han91}.

\begin{figure}
\psfig{file=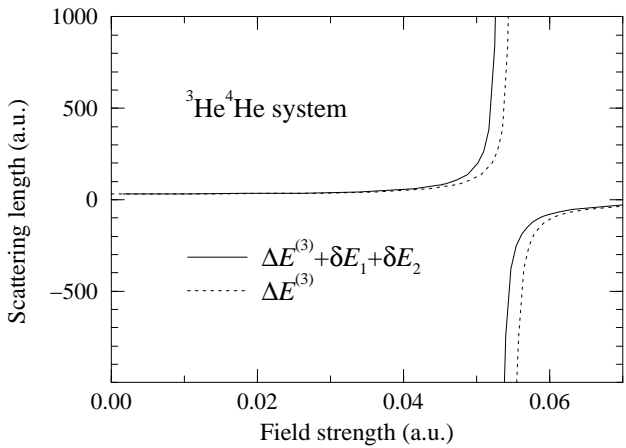,width=3.3in,%
bbllx=10mm,bblly=194mm,bburx=96mm,bbury=256mm}
\caption{The scattering length for the atomic $^3$He$^4$He system as
function of the strength of the external electric field.  The free-space
interaction is LM2M2 \protect\cite{azi93}. The induced interaction is
from Eq.(\ref{dv}) where the solid curves include all terms and the
dotted curves only include the first term.}
\label{fig1}
\end{figure}

Although the effective two-body interaction in Eq.(\ref{dv}) breaks the
rotational symmetry, the decisive $s$-wave scattering length can still
be calculated as function of the strength of the field.  The result for
$^3$He$^4$He is shown in Fig. \ref{fig1}, where an infinite scattering
length appears at $\cal E$ = 0.053 a.u.  Thus an $s$-state at zero
energy is produced by this external field at this point, i.e. the Efimov
conditions are fulfilled for the three-body system $^4$He$^3$He$_2$. In
Fig.\ref{fig1} we also see that the last term in Eq.(\ref{dv}) is rather
small for the present set of parameters. Other systems with larger
polarizability may produce a zero energy $s$-state for substantially
smaller fields.

For the $^3$He$_2$ system the scattering length is roughly
constant ($\approx$ 14 a.u.) over the range of the field strength in
Fig. \ref{fig1}. The related divergence and the bound $s$-state appears
at $\cal E$ = 0.067 a.u.  Such strong fields are possible to obtain with
todays femto-second lasers \cite{Pos96}.

Accurate computations are unfortunately rather difficult even without
an external field \cite{esr96,mot98,nie98}. However, to assess the
possibility of the effect an order of magnitude estimate should be
made before more elaborate efforts are mobilized.  The three-body
bound state energies can be obtained with a smaller numerical effort
and a relative accuracy better than 50\% by use of simple attractive
gaussian two-body interactions with the correct $s$-wave scattering
lengths \cite{nie98}.  We choose realistic range parameters and for a
given strength of external field adjust the depth of the gaussian to
reproduce the scattering lengths shown in Fig. \ref{fig1} for
$^3$He$^4$He. The accuracy of the resulting three-body binding
energies relative to the strength of the potential is around
$10^{-3}$.

\begin{figure}
\psfig{file=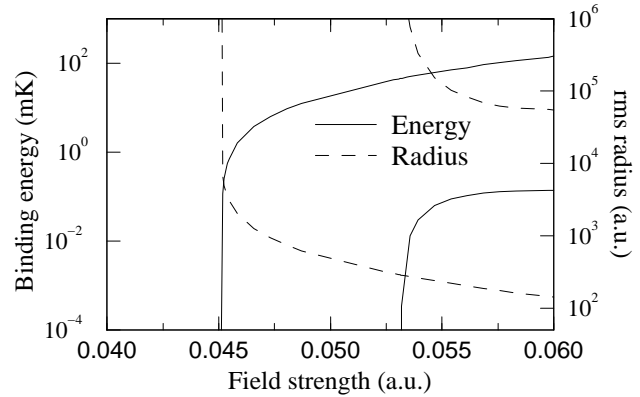,width=3.3in,%
bbllx=12mm,bblly=202mm,bburx=96mm,bbury=256mm}
\caption{The binding energies (solid curves) and root mean square
radii (dashed curves) for the ground and first excited states of
$^4$He$^3$He$_2$ as functions of the strength of the external
field. The two-body interactions are $V_0({\cal E})$ exp($-r^2/b^2$),
where $b=$ 10.546 a.u. for $^4$He$^3$He and 11.265 a.u. for
$^3$He$^3$He. The strengths $V_0({\cal E})$ produce the same $s$-wave
scattering length as the LM2M2 potential with the field ${\cal E}$.}
\label{fig2}
\end{figure}

The resulting calculated sizes and binding energies of the ground and
first excited states for $^3$He$^4$He$_2$ are shown in Fig. \ref{fig2}
as functions of the strength of the external field.  Below $\cal E$=
0.045 a.u. the three-body system is unbound.  Above this three-body
threshold the ground state binding energy increases quickly to a level of
about 0.1~K.  Just below the two-body $^3$He$^4$He threshold ($\cal E$=
0.053~a.u.) the first excited state appears with a binding energy quickly
increasing to about 1 mK.  The root mean square radii correspondingly
decrease from infinity at the thresholds to about 10 a.u. and 200 a.u.,
respectively.  Infinitely many bound three-body states with behavior
similar to these lowest states must appear at the two-body threshold.

An electric field of 0.053 a.u.=2.7 V/{\AA} induces a dipole moment
of 0.074 a.u. which corresponds to the displacement of the center
of the electron cloud by approximately 0.04 a.u. from the nucleus.
This might involve a substantial change of the electronic structure of
the atom and the original Van der Waals interaction would therefore also
change. However it is not unlikely that a more accurate calculation,
consistently including these effects, would produce Efimov conditions
already for a weaker external field.

\paragraph*{Observing the effect.}

Better suited systems can probably be found. However almost inevitably
two identical particles are needed, since their identical scattering
lengths against the third particle can then be tuned simultaneously.
The two-body subsystems should all be unbound but as near as possible
to the threshold for binding. Two atoms of naturally occurring isotopes
are almost always bound in a dimolecule. More complicated molecules
therefore seem to be needed, but then the number of combinations also
become virtually infinitely large. It is in this connection interesting
that new atomic negatively charged three-body structures recently were
suggested \cite{rob97}.

Providing the Efimov conditions must be supplemented by production and
detection of these states either by their decay or by increased
scattering cross sections. The binding energies are exceedingly small
and probably beyond the sensitivity and resolution of the experimental
equipment. On the other hand the spatial extension seems to be a
directly accessible observable. A grid with holes of variable sizes is
a direct tool to determine the radius of the created three-body state
\cite{lou96}. The intensity of appropriate systems passing the grid
then changes drastically when the Efimov states constitute a
substantial part of the molecular bound states hitting the grid.

Detailed design of an experiment is beyond the scope of the present
letter. However, a sketch of an experiment could be to let a beam of a
Borromean system in its ground state pass into a region with an
external electric field.  By photon absorption the system must then be
excited into the Efimov states which would be stopped at the grid.
The Efimov effect might also manifest itself in a number of resonances
in a three-body system which is slightly off the two-body threshold.
Scattering experiments could then provide the decisive detection
signals.

\paragraph*{Conclusion.}
Weakly bound three-body systems interacting via short-range two-body
potentials may exhibit spectacular properties exemplified by Borromean
systems and the Efimov effect. The latter occurs when the scattering
lengths are sufficiently large, or equivalently the virtual or bound
two-body $s$-states are sufficiently close to zero. A number of
excited states would then appear with very small binding energies and
correspondingly large spatial extensions. The occurrence conditions
for these Efimov states are rather sharply defined and even the lowest
of these peculiar states is difficult to localize. 

It would be a tremendous advantage if the appropriate two-body
interactions could be adjusted to meet the necessary conditions. We
propose an external electric field as the vehicle for fine tuning the
effective two-body interaction to produce a zero energy $s$-state.
The three-body computation with this field dependent two-body
interaction then reveals the Efimov states at a particular field
strength. As an illustration we used the atomic helium trimer
systems. We estimated the field strength necessary to reach the
occurrence conditions for the Efimov effect and found it on the limit
of present day technology. However, it is conceivable that other
systems are closer to the threshold and therefore are better suited
candidates.  Appropriate realistic systems should be searched for and
investigated.  We shall here be content with the demonstration that
the idea of controlled creation of the occurrence conditions for the
spectacular Efimov states is entirely feasible.

\paragraph*{\bf Acknowledgments.} We thank K.~M{\o}lmer and T.~Andersen
for helpful discussions.


\begin{thebibliography}{99}
\bibitem{efi70} V.~Efimov, Phys.~Lett.~{\bf B33}, 563 (1970);
Sov. J. Nucl. Phys. {\bf 12}, 589 (1971).

\bibitem{fed93} D.V.~Fedorov and A.S.~Jensen, Phys.~Rev.~Lett. 
{\bf 71}, 4103 (1993).

\bibitem{tho35} L.H. Thomas, Phys. Rev. {\bf 47}, 903 (1935).

\bibitem{fed94} D.V.~Fedorov, A.S.~Jensen and K.~Riisager, Phys.~Rev.~Lett.
 {\bf 73}, 2817 (1994). 

\bibitem{esr96} B.D. Esry, C.D. Lin and C.H. Greene, 
Phys.~Rev.~{\bf A54}, 394 (1996).

\bibitem{mot98} A.K. Motovilov, S.A. Sofianos and E.A. Kolganova,
J. Phys. {\bf B31}, 1279 (1998).

\bibitem{nie98} E.~Nielsen, D.V.~Fedorov and A.S.~Jensen, 
J. Phys. {\bf B31}, 4085 (1998).
 
\bibitem{vog97} J.M. Vogels, C.C. Tsai, R.S. Freeland, S.J.J.M.F Kokkelmans, 
B.J. Verhaar and D.J. Heinzen, Phys. Rev. {\bf A56}, R1067 (1998).

\bibitem{cou98} Ph. Courteille, R.S. Freeland, D.J. Heinzen, 
F.A. van Abeelen and B.J. Verhaar, Phys. Rev. Lett. {\bf 81}, 69 (1998).

\bibitem{ino98} S. Inouye, M.R. Andrews, J. Stenger, H.-J. Miesner, 
D.M. Stamper-Kurn and W. Ketterle,
Nature (London) {\bf 392}, 151 (1998).


\bibitem{han91} {\it CRC Handbook of Chemistry and Physics}, 72nd edition, 
CRC Press 1991, p. {\bf 10}-194.

\bibitem{ket92} W. Ketterle and D.E. Pritchard, Applied Physics 
{\bf B54}, 403 (1992).

\bibitem{azi93} R.A.~Aziz and M.J.~Slaman, J.~Chem.~Phys. {\bf 94}, 8047
(1993).

\bibitem{Pos96} J.H.~Posthumus, A.J.~Giles, M.R.~Thompson and K.~Codling,
J. Phys. {\bf B29}, 5811 (1996).

\bibitem{rob97} F. Robicheaux, Preprint, APS E-print database (1997),
aps1997jul01001

\bibitem{lou96} F.~Lou, C.F.~Giese, and W.R.~Gentry, J.~Chem.~Phys
{\bf 104}, 1151 (1996).

\end{thebibliography}
\end{document}